\documentclass{preprint}

\usepackage{graphicx}% Include figure files

\usepackage{times}
\usepackage{amsmath}
\usepackage{amssymb}
\usepackage{bm}
\usepackage{physics}
\usepackage{color}
\usepackage{upgreek}

\title{Zero-magnetic-field Hall effects in artificially corrugated bilayer graphene}

\author{Sheng-Chin Ho$^{1\dagger}$, Ching-Hao Chang$^{1,2\dagger}$, Yu-Chiang Hsieh$^{1}$, Shun-Tsung Lo$^{1}$, Botsz Huang$^{1}$, Thi-Hai-Yen Vu$^{1}$, Carmine Ortix$^{3,4}$, Tse-Ming Chen$^{1,2\ast}$}

\begin{document}

\maketitle

\begin{affiliations}
 \item Department of Physics, National Cheng Kung University, Tainan 701, Taiwan
 \item Center for Quantum Frontiers of Research \& Technology (QFort), National Cheng Kung University, Tainan 701, Taiwan
 \item Institute for Theoretical Physics, Center for Extreme Matter and Emergent Phenomena, Utrecht University, Princetonplein 5, NL-3584 CC Utrecht, Netherlands
 \item Dipartimento di Fisica ``E. R. Caianiello'', Universit\'a di Salerno, IT-84084 Fisciano, Italy
\end{affiliations}

\noindent $^\dagger$ These authors contributed equally to this work.\\
\noindent $^\ast$To whom correspondence should be addressed; E-mail:  tmchen@phys.ncku.edu.tw.\\

\begin{abstract}

The ability to engineer the electronic band structure and, more strikingly, to access new exotic phase of matter has been the cornerstone of the advance of science and technology. Twisting van der Waals materials to form moir\'e superlattice is a powerful paradigm and can drive graphene from a normal metallic state into an insulating, superconducting, or ferromagnetic states. Here, we present a new route to create non-trivial band structure and consequently an exotic phase of matter via lithographically patterned strain (lattice deformation). This method is used to realize an artificially corrugated bilayer graphene wherein the real-space and momentum-space pseudo-magnetic fields (Berry curvatures) coexist and have nontrivial properties, namely, the Berry curvature dipole. This new class of condensed-matter systems enables us to observe the so-called nonlinear anomalous Hall effect and a new type of Hall effect without breaking the time-reversal symmetry. Such artificial material system and our approach to unconventional electronic states may open an avenue of geometrical and/or topological quantum phenomena as well as that of band engineering in van der Waals crystals.

\end{abstract}

The Hall effect family has been an important driving force behind the advance of solid-state physics and technology over the past century, e.g. the quantum Hall effect opened the door to explore the world of matter via topology and defined today's metrology standard\cite{klitzing_prl80,thouless_prl82}, the anomalous Hall effect led to the discovery of the Berry curvature. The Berry curvature in momentum space -- which in condensed matter characterizes the geometric properties of electronic Bloch bands and can be viewed as a momentum-space magnetic field -- underlies a wide variety of spectacular phenomena\cite{nagaosa_rmp10,xiao_rmp10} and holds promise for future technologies such as valleytronics\cite{schaibley_nrm16,mak_science14,gorbachev_science14,sui_np15,shimazaki_np15}. For a solid to possess a nonzero Berry curvature, either the time-reversal or the spatial inversion symmetry has to be broken. This is typically done by coupling an external magnetic field or electrical displacement field to explicitly break the time-reversal or inversion symmetry, respectively; or, even more simply, can be realized in magnetic or noncentrosymmetric materials where the time-reversal or inversion symmetry has been spontaneously broken. Recently, it has been suggested that when the system has very low crystalline symmetry the Berry curvature can exhibit nontrivial properties, the so-called Berry curvature dipole, typically resulting from tilted Dirac or Weyl cones. This can reveal striking phenomena\cite{low_nm17,xu_np18}, including the nonlinear anomalous Hall effect (AHE)\cite{moore_prl10,sodemann_prl15,zhang_arxiv18,shi_arxiv18,ma_arxiv18,kang_arxiv18}. The observation of the nonlinear AHE in topological semimetal WTe$_2$ has thus attracted immense attention\cite{ma_arxiv18,kang_arxiv18} and shattered the general belief that broken time-reversal symmetry is a necessary condition to have a nonvanishing Hall conductivity.

On the other side, the real-space pseudo-magnetic field in two-dimensional materials like graphene is also of significant interest because it provides a means to modify the band structures and consequently the electronic properties through mechanical deformations or, more precisely, by engineering the lattice strain\cite{McCann_prl06,guinea_np10,low_nl10,mucha_prb11,tang_np14,levy_science10,jiang_nl17,liu_nn18}. It was found that a small lattice deformation can create a very large pseudo-magnetic field on the order of tens or hundreds of tesla\cite{guinea_np10,levy_science10}. This pseudo-magnetic field can lead to pseudo-Landau levels (pseudo-LLs) and a quantum valley Hall effect analogue to the quantum spin Hall effect in topological insulators simply because the pseudo-magnetic field has opposite signs for graphene's two valleys (as demanded by time-reversal symmetry), and therefore gives rise to counter-propagating skipping orbits and edge states. Despite being striking, it had remained a challenge to expand such a pseudo-magnetic field from a local nanoscale bubble or ripple that can only be detected using scanning tunneling microscope\cite{levy_science10,jiang_nl17} to a large mesoscopic system for transport measurements until very recently\cite{liu_nn18}. More importantly, most research so far has focused on creating a nearly uniform\cite{guinea_np10,low_nl10,levy_science10,jiang_nl17} or an oscillating\cite{liu_nn18} peseudo-magnetic field on monolayer graphene (MLG), which does not much lower the spatial symmetry.

Here, we create a new class of condensed-matter system in which both the Berry curvature dipole and real-space pseudo-magnetic field dominate, and furthermore use this system to demonstrate the nonlinear AHE and a new type of a linear response Hall effect, which may be viewed as a cousin of the Magnus Hall effect that was recently predicted\cite{papaj_prl19}, without breaking the time-reversal symmetry. Such a system is realized by placing bilayer graphene (BLG) onto an artificially corrugated hexagonal boron nitride (hBN) substrate to simultaneously break the inversion symmetry and induce pseudo-magnetic fields through mechanical deformations, as shown schematically in Fig.~1a. There are two distinct features that make our design unique from others\cite{levy_science10,jiang_nl17,liu_nn18}. First, the use of BLG instead of MLG breaks the inversion symmetry in the lattice-deformed (or strained) regions. In this case, the interlayer coupling is expected to be modulated in accordance with the lithographically-patterned lattice deformation (strain) and play a significant role in the band structure modulation, conceptually similar to the twisted bilayer graphene wherein the interlayer coupling is instead modulated by the moir\'e pattern to drive graphene from a normal metallic state into an insulating\cite{cao_nature18a}, superconducting\cite{cao_nature18b}, or ferromagnetic stats\cite{sharpe_science19}. Second, the corrugated steps are designed to be further apart ($100$~nm), spatially separating the strain field and hence the antiferromagnetic pair of pseudo-magnetic fields $B_\text{S}$ from each other (Fig.~1b). This is distinct from a continuous sinusoidal-like $B_\text{S}$ created in a continuously corrugated system\cite{liu_nn18}, in which the first-order moment of the pseudo-magnetic field is zero. In contrast, spatially separated antiferromagnetic pairs of $B_\text{S}$ will have a nonzero first-order moment for each of the K and K$'$ valley, analogous to the Berry curvature dipole in momentum space\cite{sodemann_prl15}, and remarkably gives rise to a Rashba-like valley-orbit coupling (which will be discussed in more detail below).

The best and easiest way to understand and classify a condensed-matter system is through the band structure. We, therefore, construct an effective two-band Hamiltonian to calculate the low-energy band structure of our corrugated BLG system\cite{McCann_prl06}. The effect of lattice deformation is cast in terms of a periodic inhomogeneous pseudogauge field $\mathbf{A}=A_y(x) \hat{j}$, which results in a pseudo-magnetic field $\mathbf{B}_{\text{S}}= (\partial_x A_y) \hat{k}$. We further include the skew interlayer coupling that crucially warps the iso-energetic lines in strained BLG systems\cite{mucha_prb11} (see Methods). Figure~1c shows one of the tilted mini-Dirac cones that are theoretically predicted in our corrugated BLG. Such a mini-Dirac cone is created via the crossing and anti-crossing of the pseudo-LLs and the interlayer coupling within the BLG. The mini-Dirac cone is tilted and has non-trivial and anisotropic energy dispersion. The Berry curvature in this tilted mini-Dirac cone has been calculated and exhibits a non-uniform distribution, which leads to a nonzero Berry curvature dipole (see Fig.~1g \& Extended Data Fig.~2). Figures~1d-g schematically show the Fermi circles and Berry curvatures for a series of different classes of condensed-matter systems. We note that the artificially corrugated BLG has a very unique and exotic band dispersion and Berry curvature, in which the three-fold rotational symmetry and both the mirror lines can be broken.

To fabricate the devices (Fig.~1h), we first prepare the corrugated hBN substrate by shaping the surface of a mechanically exfoliated hBN flake using the electron-beam lithography and plasma etching to create the desired surface topography in selected regions (highlighted by a red-dashed rectangle in Fig.~1h). A suitable exfoliated BLG flake is placed onto this corrugated hBN substrate using a polymer-based dry transfer technique, followed by thermal anneal in vacuum, to realize this artificially corrugated heterostructure. Standard nanofabrication processes are then used to create our multi-terminal device with Hall-bar geometry (Fig.~1h; see Methods \& Extended Data Fig.~1 for the details of the whole fabrication procedure). Following the device fabrication, we use atomic force microscope (AFM) to examine the surface topography. The AFM topography image (Fig.~1i) shows that the BLG film is nicely deformed into a corrugated pattern.

We now move on to report the observation of nonvanishing nonlinear and linear transverse (Hall) responses without breaking the time-reversal symmetry -- hereafter referred to as the nonlinear AHE and the pseudo-magnetic-field induced planar Hall effect (pseudo-PHE), respectively. It has generally been believed that the broken time-reversal symmetry is a necessary condition to have a nonvanishing Hall conductivity (or resistivity). It is because under time-reversal operation any form of the magnetic field (regardless of being in real- or momentum-spaces) will have opposite signs for carriers with opposite magnetic quantum numbers (e.g. valleys, spins), and therefore the overall Hall contributions are cancelled out in most cases. Nonlinear AHE has been predicted to be the first exception as the second-order nonlinear Hall contributions from the anomalous velocity do not cancel out when there is a Berry curvature dipole\cite{sodemann_prl15}, which typically results from tilted Dirac or Weyl cones. Such nonlinear AHE was predicted to occur in some emergent materials with very low crystalline symmetry\cite{sodemann_prl15,zhang_arxiv18,shi_arxiv18} and has recently been observed in 1T$_d$~WTe$_2$\cite{ma_arxiv18,kang_arxiv18}.

Notably, the nonlinear AHE is also observed in our corrugated BLG, an artificial material system that has many pseudo-LL-induced tilted mini-Dirac cones (Figs.~1c, g), and the Berry curvature dipole on these tilted mini-Dirac cones can be theoretically predicted (Extended Data Fig.~2). To observe the nonlinear AHE in experiments, we utilize the Hall measurement setup but record the nonlinear response to the driving current $I_x (\omega)$, i.e. the response at the second-harmonic frequency $2\omega$. Figure~2a shows the nonlinear Hall voltage $V^{2\omega}$ in our corrugated BLG at zero magnetic field. The nonlinear Hall voltage observed here exhibits a complicated structure as a function of $V_g$, which is not unexpected for our corrugated BLG with a complicated band structure (see Fig.~4). The Berry curvature is known to be very sensitive to the Bloch wavefunctions (geometric properties of the band structures) and thus the AHE commonly has a fluctuant Hall resistivity with respect to Fermi energy\cite{fang_science03}. Besides, the dipole moment changes sign whenever the chemical potential passes through a mini-Dirac cone band gap, and therefore will fluctuate significantly with successive sign changes as a function of chemical potential. It is important to stress that the fluctuation details of the nonlinear AHE is highly reproducible and robust during thermal cycling, eliminating the possibility that this response is due to disordered scattering and/or interference effects.

The nonlinear AHE based on the Berry curvature dipole scenario has two distinct features: (i) the Hall voltage will depend quadratically on the driving current, and (ii) the longitudinal signal is absent, which ideally will give rise to a $90$-degree Hall angle. We demonstrate these two features in Fig.~2b to verify that our observed nonlinear Hall effect is indeed due to the Berry curvature dipole. The measured nonlinear AHE voltage $V^{2\omega}_y$ (red squares) is shown to have a square relation with the driving current, whereas for comparison the linear response Hall voltage $V^{\omega}_y$ (blue circles; pseudo-PHE that will be discussed below) scales linearly with the current. We also find that the longitudinal voltage measured in the nonlinear response regime $V^{2\omega}_x$ (black triangles) is indeed zero and does not respond to the current increase. Furthermore, we demonstrate that the nonlinear AHE observed in our corrugated BLG is always along the transverse direction with a $90$-degree Hall angle no matter whether the current is applied either perpendicular or parallel to the corrugation direction (see Extended Data Fig.~3). This provides important evidence to further verify the origin of our observed nonlinear AHE since one of the most distinct characteristics of the nonlinear AHE originating from the Berry curvature dipole is that the nonlinear response must be perpendicular to the directions of the current and the Berry curvature (i.e., in the $z$ direction), similar to the directional dependence of the typical Hall effect. It rules out the possibility that the observed nonlinear AHE is driven by the skew scattering due to the inherent chirality of itinerant electrons\cite{Isobe_arXiv19}, another new interesting mechanism that may also induce nonlinear transport under time-reversal symmetry, because the nonlinear transport of the inherent chirality origin would give rise to a nonlinear response in both transverse (Hall) and longitudinal directions. These unique characteristics also distinguish the nonlinear AHE of Berry curvature dipole origin from other nonlinear transport in noncentrosymmtric van der Waals materials under broken time reversal symmetry\cite{krstic_jcp02,pop_ncomm14,qin_ncomm17,ideue_np17}.

The absence of the nonlinear AHE in both simple BLG and corrugated MLG (Extended Data Fig.~4) has further supported our scenario. In addition, the temperature dependence of the nonlinear AHE has been investigated by both experimental and theoretical approaches (Fig.~2c). The nonlinear AHE is determined by the Berry curvature dipole over the occupied states and hence the effect of increasing temperature on the nonlinear Hall voltage is based on the Fermi-Dirac distribution function if the electron transport relaxation time is constant with temperature changes\cite{sodemann_prl15} (see Methods). There is a good agreement between experiment and theory except at the high temperature regime, which may be due to the decrease of the relaxation time at high temperature since in our theoretical calculation, for simplicity, we assume the relaxation time to be constant. Note that the temperature dependence of the nonlinear AHE (and also its fluctuating behaviour with $E_F$) observed in our corrugated BLG is in conflict with other known possible scenarios that may induce the nonlinear AHE\cite{Isobe_arXiv19}. Although the results and the agreement between our experiments and the theory based on the Berry curvature dipole scenario provide convincing evidence pointing to this as the origin of the nonlinear AHE in our corrugated BLG, we have no intention to conclude that the Berry curvature dipole is the only possible mechanism as the system that hosts both the real-space and momentum-space pseudo-magnetic fields (Berry curvatures) may create other new mechanisms beyond our knowledge.

The Berry curvature dipole lies at the root of the nonlinear AHE. Hence, in order to understand it in our corrugated BLG, we derive the Berry curvature dipole (Fig.~2d) from the measured nonlinear Hall voltage (Fig.~2a) and other transport properties (see Methods for the detailed derivation). One of the most interesting features of the Berry curvature dipole in our corrugated BLG is that the value is relatively small near the charge neutrality point but become large when the Fermi energy is tuned away to both the electron and hole sides. It is likely because that the band dispersion is relatively symmetric near the NP, in which the mini-Dirac cones are less titled and consequently the Berry curvature dipole is expected to be smaller (see Fig.~4 below for the band structure predicted for the corrugated BLG). For comparison, we have performed the theoretical calculation of the Berry curvature dipole using the band structure predicted for our corrugated BLG, and the profile as shown in Fig.~2e is qualitatively consistent especially on the hole side. This nontrivial energy dependence of the Berry curvature dipole provides new insights and access into the interplay between the momentum- and real-space pseudo-magnetic fields. It is important to stress that the magnitude of the Berry curvature dipole that is theoretically calculated from the band structure of the corrugated BLG is close to the experimentally measured value, and, more importantly, is about two and one order of magnitude larger than those obtained in monolayer\cite{xu_np18} and bilayer\cite{ma_arxiv18} WTe$_2$, respectively.

We notice that the numbers of the peaks of the Berry curvature dipole are different between experiment and theory. This may be due to the fact that a relatively ideal and simple model is used in our theoretical calculations. We neglect electron-hole asymmetry, next-nearest-neighbor hopping processes, and interlayer inhomogeneous shear strains. It is known that the inhomogeneity and a large variety of symmetry breaking can break degeneracy and open band gap so as to induce Berry curvature. Therefore, more mini-Dirac cones in a more complicated band structure formed by warped pseudo-LL magnetic states are expected. This could lead to a more fluctuant structure with respect to Fermi energy since the Berry curvature is very sensitive to the geometric properties of the band structures\cite{fang_science03}.

Next, we move on to the nonvanishing transverse response in the linear regime, i.e. the pseudo-PHE. Figure~3a shows the unexpected linear transverse (Hall-like) response, with an unusual back-gate dependence, observed in our corrugated BLG at zero magnetic field. To understand its origin, we start by considering the physical velocity of the electrons
\vspace*{1mm}
\begin{equation}
\bm{v}(\mathbf{k}) = \frac{1}{\hbar} \frac{\partial \epsilon(\mathbf{k})}{\partial \mathbf{k}} + \dot{\mathbf{k}} \times \mathbf{\Omega}(\mathbf{k}),
\vspace*{1mm}
\end{equation} 
where the first term denotes the group velocity from the band dispersion $\epsilon(\bm{k})$ and the second term is the anomalous velocity arising from the Berry curvature $\mathbf{\Omega}$. Note that the second term is analogous to the Lorentz force in momentum space and therefore $\mathbf{\Omega}$ is commonly viewed as a momentum-space magnetic field. Since the K and K$'$ valleys have opposite $\mathbf{\Omega}$ as required by the time-reversal symmetry $\mathbf{\Omega}(\mathbf{k})=-\mathbf{\Omega}(-\mathbf{k})$, the linear transverse response from anomalous velocities for the two valleys cancels out. This is why a time-reversal breaking field is generally required for the Hall effect related to the Berry curvature to be observed\cite{mak_science14}. However, the transverse contribution from the band structure can be non-zero if $\int d^2 \mathbf{k} \delta(\epsilon(\mathbf{k})-\epsilon_{\text{F}}) \partial_{x} \epsilon(\mathbf{k}) \partial_{y} \epsilon(\mathbf{k}) \neq 0$, where $\epsilon_{\text{F}}$ is the Fermi energy and $\partial_{x,y}$ denotes $\partial / \partial k_{x,y}$ (see Methods). In other words, a linear Hall-like transverse response can appear if one can find a time-reversal invariant pseudo-magnetic field to distort the band dispersion and Fermi circle into anisotropy. This is similar to the planar Hall effect originating from an in-plane magnetic field\cite{kandala_ncomm15,taskin_ncomm17} or the electronic nematicity\cite{wu_nature17} so as to cause resistance anisotropy. Here, however, it is the out-of-plane pseudo-magnetic field together with its interplay with the interlayer coupling that are responsible for the nontrivial anisotropy of the Fermi circle. The observed Hall-like response can therefore be viewed as a pseudo-magnetoresistance anisotropy or pseudo-magnetic-field induced PHE (pseudo-PHE).

Figure~3b shows the transverse resistivity calculated using the theoretically predicted band dispersion of our corrugated BLG devices. There is a good qualitative and quantitative agreement between the experimental observations and the theoretical predictions. We furthermore present the temperature dependence of the pseudo-PHE in Figs.~3c (experiment) \& 3d (theory). As temperature decreases, resistivity fluctuations (or oscillations) emerge, superimposed on the two arms of the pseudo-PHE (i.e., on the electron and hole sides). Note that although the two arms are very robust to temperature variations, the pseudo-PHE near the charge neutrality point (NP) and the resistivity fluctuations are rather sensitive as these fine features indicate the strong band distortion near the Dirac point. The unusual gate-dependence of the pseudo-planar Hall contribution intrigues us to further investigate the angular variation of the anisotropy axis of the pseudo-magnetoresistance (see Fig.~3e). The angle $\theta$ between the principle axes of the anisotropic resistance and the current direction can be estimated by considering a rotation matrix to diagonalize the measured resistivity tensor (see Methods). Theoretical calculation of $\theta$ based on the predicted band structure of the corrugated BLG has also been presented, and is in good agreement with the experimental results. Strikingly, the anisotropy axis has been found to change direction as a function of gate voltage (i.e., Fermi energy) while the direction of the pseudo-magnetic field $B_{\text{S}}$ is fixed. This makes the pseudo-PHE unique from most earlier reported PHE and magnetic anisotropy\cite{kandala_ncomm15,taskin_ncomm17} since the anisotropy axis is generally set by the magnetic field and thus would not change direction when the field is fixed. This finding might have important implication in both fundamental research and technological applications as it provides an alternative means to rotate the resistance anisotropy on par with varying the doping levels\cite{wu_nature17}. 

In addition, it may be worth stressing that the current is applied perpendicular to the corrugation. It means that the current is on the principle axes of the magnetic anisotropy and therefore no planar Hall contribution should have appeared if there were not interlayer interaction to warp the band dispersion and tilt the resistance anisotropy away from the corrugation direction (see Fig.~1c, g and later Fig.~4 for further discussion). In order to confirm that the pseudo-PHE is a consequence of the interplay between the pseudo-magnetic field and interlayer interaction, we have performed the same experiments on simple BLG and corrugated MLG (which has pseudo-magnetic fields but no interlayer interaction), and there are no clear signs of the pseudo-PHE in these two systems (Extended Data Fig.~5). Note that the pseudo-PHE refers to only the nonzero Hall voltage that can exist over a wide range of gate voltage (Fermi energy) and has the two-arm structure. In contrast, consistent results of the pseudo-PHE are reproduced in multiple devices and for different cooldowns (Extended Data Fig.~6), ruling out the possibility that the Hall response is due to disordered scattering or sample inhomogeneity. We have also conducted magnetic-field dependence to demonstrate that our system is indeed under time-reversal symmetry (Extended Data Fig.~7), ruling out the possibility that the observed Hall signal is due to the magnetic contamination or other time reversal symmetry breaking fields (see Methods).

Having established the existence of nonlinear AHE and pseudo-PHE, we now move to explain, from a microscopic viewpoint, why our BLG corrugation can lead to exotic band structure and Berry curvature. The pseudo-magnetic states will form as a result of the strain-induced pseudo-magnetic field, as shown in the calculated band structure in Figs.~4a \& b. Electron transport in the vicinity of the BLG corrugation are valley-dependent because electrons in different valleys experience different pseudo-magnetic fields (Fig.~1b). For instance, the low-energy electronic states in both K and K$'$ valleys are snake-like magnetic states trapped around the corrugation, but move in opposite directions since the effective Lorentz force they experience is opposite, illustrated in Fig.~4c. This valley-orbit coupled nature is general for all electronic states in the BLG and leads to a Rashba-like valley-orbit coupling -- the band structures in K$'$ and K not only shift but also slightly slope towards opposite momentum (Fig.~4a). However, the band structure along the $k_x$ axis has a conventional symmetric Brillouin zone due to the periodicity along the $x$-axis in our BLG corrugation (Fig.~4b). Note that this Rashba-like valley-orbit coupling is also expected to occur in MLG corrugation (Extended Data Fig.~8) since it is only due to nonzero first-order moment for each of the K and K$'$ valley and their associated snake-like magnetic states and has nothing to do with the interlayer coupling.

The interlayer coupling is expected to play a crucial role in our corrugated BLG because the lattice deformation will significantly modulate the interlayer skew hopping and hybridization. When a uniaxial strain (lattice deformation) is applied to bilayer graphene via our corrugation structure, it breaks the hexagonal lattice symmetry and changes the interlayer hopping by making them direction dependent, because the relative position and distance between the hopping sites have all changed and become different along different directions. It is apparent that such direction-dependent hopping will lead to an anisotropic transport and hence distort the electron dispersion\cite{mucha_prb11}. Figures~4d \& e show that the skew interlayer hopping not only hybridizes and mixes the pseudo-magnetic states but also distort the band dispersions. The energetic contour of these pseudo-magnetic states is therefore warped, and several band crossings are also induced to form tilted mini-Dirac cones that host large Berry curvature dipoles (see Figs.~4d-f). The warped valley-orbit coupled band dispersion and the Berry curvature dipoles hosted by the titled mini-Dirac cones, respectively, result in the pseudo-PHE and nonlinear AHE that coexist in our corrugated BLG system. For comparison, we also perform theoretical calculations in a corrugated MLG with the same geometry, and as expected no pseudo-PHE and nonlinear AHE exhibit in this MLG corrugation, same as the experimental results (see Methods).

Last but not least, we re-establish the pseudo-PHE by using a separate theoretical approach that do not rely on the electronic band structure. A simple effective model based on solving electron dynamics is employed to prove that a nonzero Hall response will result from the coexistence of pseudo-magnetic field and effective electric fields that are created by the skew interlayer coupling, and can exist in both multi- and single-corrugation BLG systems (see Methods). The consistency of results from different theoretical approaches indicates that the pseudo-PHE connects to electron motions driven by both effective fields and thus is robust against the temperature increase (Fig.~3d). We notice that the pesudo-PHE reported here may be viewed as a cousin of the the recently proposed Magnus Hall effect\cite{papaj_prl19}. The Magnus Hall effect is induced by non-uniform electric potential created by corrugated local gates, which is analogue to our system where non-uniform pseudogauge field (i.e. pseudo-magnetic vector potential) is created by corrugated strains.

In short, we demonstrate an approach to engineering the band structure to access exotic phases of matter, complementary to twistronics\cite{cao_nature18a,cao_nature18b,sharpe_science19} or pressuring the materials\cite{yankowitz_nature18}. We employ this technique to drive bilayer graphene into an unconventional electronic state with tilted, distorted mini-Dirac cones, which enables us to demonstrate two types of Hall effects under time-reversal symmetry. The pseudo-PHE (or the pseudo-magnetoresistance anisotropy) is a consequence of nontrivial, anisotropic band distortion induced by the interplay of strain-modified intralayer (manifested in pseudo-magnetic fields) and interlayer couplings. On the other hand, the observation of the nonlinear AHE indicates that it is possible to artificially introduce some of the most striking physical quantities -- here, the Berry curvature dipole -- in a conventional well-known system via simple modifications. The capability to create a platform with both the real-space and momentum-space pseudo-magnetic fields (Berry curvatures), together with the ease of fabrication and up-scaling with our approach, may open lots of possibilities for research in geometrical and/or topological quantum phenomena and their technological applications.

\section*{Acknowledgements} 
\vspace{-5mm}
We thank T.-R.~Chang, Y.-C.~Chen, C.-J.~Chung, S.-Z. Ho, Y.-D. Liou, L.~W.~Smith and J.~I-J.~Wang for helpful discussions and/or technical assistance. This work was supported by the Ministry of Science and Technology (Taiwan), and the Headquarters of University Advancement at the National Cheng Kung University.

\section*{Author Contributions}
\vspace{-5mm}
S.-C.H. performed the measurements and analysed the data with help from S.-T.L. and Y.-C.H. C.-H.C. developed the theory and performed the calculations with assistance from B.H., C.O. and T.-M.C. S.-C.H., Y.-C.H. and T.-H.-Y.V. developed and performed the sample fabrication. S.-C.H. and T.-M.C. designed the experiment. S.-C.H., C.-H.C. and T.-M.C wrote the manuscript with input from all authors. T.-M.C. supervised the project.

\section*{\large METHODS}
\vspace{-5mm}
\subsection{Device fabrication.}
To fabricate the corrugated bilayer graphene (BLG) devices, we use the polymer-based dry pick-up and transfer technique to assemble the hexagonal boron nitride (hBN) and BLG by van der Waals force. The whole process is shown in Extended Data Fig.~1. BLG and hBN candidates are prepared on different SiO$_{2}$ substrates with plasma-cleaning by reactive ion etching. The selected hBN is transferred to the host substrate (Extended Data Fig.~1a) and subsequently patterned into corrugation by electron-beam lithography and inductively-coupled plasma etching in an O$_{2}$, Ar and CHF$_{3}$ environment (Extended Data Fig.~1b). The bilayer graphene is afterward transferred to this corrugated hBN substrate (Extended Data Fig.~1c). After the transfer process, the whole sample is annealed in the vacuum ($<10^{-4}$~mbar) with the temperature of $350^{\circ}$C for 16 hours. The system is gradually increased from room temperature to 350 $^{\circ}$C for 7 hours, maintained at the highest temperature for 2 hours and then decreases back to the room temperature for another 7 hours. We then etch this corrugated heterostructure into a multi-terminal Hall bar geometry (Extended Data Fig.~1d) and finally form metallic contacts using the electron-beam evaporator to deposit Cu/Au with the thickness of 1.5/60~nm  (Extended Data Fig.~1e). The mean free path estimated using the temperature dependence of the resistivity ranges from $130$~nm to $500$~nm for our measured corrugated BLG devices, all of which give consistent results.

\subsection{Measurements.}
Standard four-terminal measurements are performed in a cryostat down to 1.5~K. As illustrated in Fig.~1h in the main article, an AC current of $100$~nA at $ \omega= 77$~Hz is applied to the devices unless otherwise stated, and the four-terminal longitudinal and transverse voltage at the first- ($ \omega $) and second- ($2 \omega $) harmonic responses to the excitation current are measured simultaneously using the lock-in technique (with Stanford SR830). The signal amplitude (R), in-phase component (X) and the signal phase ($\uptheta$) are all recorded during the measurements. The phases of the first- and second-harmonic responses are $0^{\circ}$ and $\pm90^{\circ}$, respectively, in agreement with the expectations for the linear and nonlinear measurements.

\subsection{Time reversal invariant pseudo-PHE}
In the main text, this pseudo-PHE has been demonstrated when the current is applied along the $x$-axis -- i.e. perpendicular to the corrugation -- to give the non-vanishing $\rho_{yx}$. In order to verify whether or not the observed effect indeed possesses time-reversal symmetry, we also perform the measurements with the current applied along the $y$-axis -- i.e. parallel to the corrugation -- to measure the $\rho_{xy}$ in the same device such to examine whether the resistivity is a symmetric tensor. Despite significant differences in the mesoscopic transport along the current direction between these two measurement setups, we find that the two off-diagonal resistivities $\rho_{xy}$ and $\rho_{yx}$ are nearly the same (Extended Data Fig.~\ref{Fig3s}a). This clearly demonstrates that the time-reversal symmetry is preserved in our corrugated BLG system; hence, the pseudo-PHE is not linked to any magnetic contamination or other time reversal symmetry breaking fields. Furthermore, we add an external magnetic field $B$ to explicitly break the time-reversal symmetry and introduce the classical Hall effect (HE) to study the interplay between them. Extended Data Figure~\ref{Fig3s}b shows that a sizable $B$ (of roughly $0.9$ and $-0.5$~T on the hole and electron sides, respectively) is needed for the Lorentz force driven HE to cancel out the Hall-like contribution from pseudo-magnetic fields. Also, since the pseudo-PHE and the HE have different symmetry (i.e. symmetric and anti-symmetric, respectively) with respect to both the external magnetic field and the sign of charge carriers, their linear combination makes the overall Hall resistivity asymmetric over $B$ and $V_g$. However, even though $\rho_{yx} (B, V_g) \neq \pm \rho_{yx} (-B, V_g)$ because of the asymmetry, we find $\rho_{yx} (B, V_g) = \rho_{xy} (-B, V_g)$ as shown in Extended Data Fig.~\ref{Fig3s}d in spite of the great differences among measurement arrangement and contact configurations (illustrated in the insets of Extended Data Figs.~\ref{Fig3s}b,c), again satisfying the Onsager relation.

\subsection{Theoretical models for bilayer graphene system with periodic corrugations}
%The nature of low-energy electrons in corrugated bilayer graphene (BLG) critically depends on the mechanical deformations that introduce not only large strain and also nontrivial interlayer coupling.
Here we introduce the derivation of an effective two-band Hamiltonian that describes the low-energy electrons in the corrugated BLG. For electronic states near the Brillouin zone corners K ($\xi =1$) and K$'$ ($\xi=-1$) in a strained BLG, the approximate two-band Hamiltonian with the $x$-axis along the corrugated (strained) direction is\cite{mucha_prb11,mcc2013} :
\begin{align}
\notag \hat{H} &=-\frac{1}{2m^\star}\left (
  \begin{array}{cc}
    0 &( \pi_0^+)^2\\
 \pi_0^2& 0  \\
  \end{array}
\right )+v_3\left (
  \begin{array}{cc}
    0 & e^{i3\phi} \pi_3 \\
e^{-i3\phi} \pi^+_3 & 0  \\
\end{array}\right )+\left (
  \begin{array}{cc}
    0 & \omega\\
\omega^* & 0  \\
\end{array}\right )+\xi \frac{u}{2} \sigma_z,\\
\notag\pi_0 &= \xi p_x+ip_y+e^{-3i\phi} A_{0}(x),\\
\pi_3 &= \xi p_x+ip_y+e^{-3i\phi} A_{3}(x),
\label{h_2band}
\end{align}
where the electron effective mass $m^\star = 0.035 m_e$ with $m_e$ being the electron rest mass, $\phi$ is the angle between the corrugated direction and the graphene zigzag direction, and the interlayer asymmetry $u = 0.8$ meV. The second term in the above Hamiltonian is contributed by the skew interlayer coupling and its velocity $v_3$ is obtained by considering  a ratio between interlayer and intralayer coupling parameters as
an interlayer-coupling strength $\gamma_1\gamma_3/\gamma^2_0 \approx 0.04$. The effective mass and coupling strengths are consistent with values estimated in experiments\cite{mcc2013}. 
The third term is contributed by the shear strain among two layers \cite{mucha_prb11} and we ignore its impact by consider $\omega = 0$. We note that the effect of onsite interlayer coupling $\gamma_1$ is already included in electron effective mass $m^\star$ \cite{mcc2013} .

We obtain Eq. (\ref{h_2band}) by transforming the xy coordinates of strained-bilayer-graphene Hamiltonian \cite{mucha_prb11} from the zigzag-armchair direction to the strain tension-contraction direction.
The pseudo-gauge fields $\mathbf{A}_0 = A_0(\cos3\phi,-\sin3\phi)$ and $\mathbf{A}_3= A_3(\cos3\phi,-\sin3\phi)$ originate from the intralayer and interlayer lattice deformations.
The intralayer pseudo-gauge field has been discussed extensively in monolayer graphene (MLG) systems.
For the out-of-plane lattice deformation induced in our corrugated substrate with a height $h(x)$, it can be written as\cite{car2016}
\begin{align}
\mathbf{A}_0=(A_{0x}, A_{0y})^{T}=\frac{\hbar\beta}{4 r_{\rm AB}}(\partial_{x}h(x))^2(\cos 3\phi,-\sin3\phi)^T,
\label{eq_a0}
\end{align} 
where the in-plane AB sublattice distance $r_{\rm AB} \approx 1.42\AA$ and the coefficient $\beta \approx 3$, predicted within a $20 \%$ range of variation\cite{set2016}. In the K valley, the gauge fields result in the pseudo-magnetic fields $\mathbf{B}_0=\nabla\times\mathbf{A}$ along the $z$ axis with magnitudes $B_0 = \partial_x A_{0y}$. In the K$'$ valley, both effective fields are the same in the magnitude, but have opposite sign because the valley index $\xi=-1$ reverses the direction of the Lorenz force. Besides, $B_0$ becomes zero when the inhomogeneous strain is along the zigzag direction (i.e. $\phi = n \times 60^{\circ}$, where $n$ is an integer)\cite{ver2015}. This indicates that the pseudo-PHE is zero when the corrugation direction is along the zigzag direction. 

Since the curvature of corrugations in our corrugated BLG is moderate, the shear strain between two layers can be neglected, and therefore the interlayer pseudo-gauge field $\mathbf{A}_3$ can be solved approximately. When a distance between the in-plane AB sublattice extends from $r_{\rm AB}$ to $r_{\rm AB}+\Delta$ due to the corrugated substrate, the interlayer AB sublattice distance in the AB-stacking BLG extends from $\sqrt{r^2_{\rm AB}+d^2}$ to $\sqrt{(r_{\rm AB}+\Delta)^2+d^2}$ with the interlayer distance $d \approx 3 \AA$. The ratio between the $|\mathbf{A}_{0}|$ and $|\mathbf{A}_{3}|$ can be obtained by comparing the changes of interlayer and intralayer lattice displacements:
\begin{align}
\notag \frac{A_{3}}{A_{0}}&=\frac{\sqrt{(r_{\rm AB}+\Delta)^2+d^2}/\sqrt{r^2_{\rm AB}+\Delta^2}-1}{(r_{\rm AB}+\Delta)/r_{\rm AB}-1}\\
 &\approx \frac{r^2_{\rm AB}}{r^2_{\rm AB}+d^2}\approx 0.18
\label{eq_a3}
\end{align}

For the pseudo-gauge fields with periodicity $A_{0}(x+{\rm L})=A_{0}(x)$, 
the exact eigenstates of the Hamiltonian in Eq.~(\ref{h_2band}) can be represented as
\begin{align}
\Psi(x,y)=e^{ik_x x} \sum_{n,\pm } a_{n,\pm}(k_x,~k_y) \times\psi_{n,\pm,k_y}(x,y),
\label{eq:eigen}
\end{align}
where $0 \le k_x < 2\pi/{\rm L}$ due to the one-dimensional Brillouin zone along $k_x$. Here $ a_{n,\pm} $ is the coefficient expended by the complete orthonormal basis 
\begin{align}
 \psi_{n,\pm,k_y}(x,y)=\frac{1}{\sqrt{2}}\left (
  \begin{array}{cc}
    1 \\
 \mp e^{2 i\phi}\\
  \end{array}
\right )e^{i(k_y y+n2\pi x/{\rm L})},
\label{eq:basis}
\end{align}
where $e^{2i\phi}=(n2\pi /{\rm L}+ik_y)/(n2\pi /{\rm L}-ik_y)$.
By direct diagonalization of the Hamiltonian Eq.~(\ref{h_2band}) on the basis $ \psi_{n,\pm,k_y}(x,y)$, we obtain the eigenstates and the corresponding band structure for periodic pseudo-gauge fields $A_{0y}$ and $A_{3y}$.  To reach an accuracy of calculated eigenenergies in Fig.~4 in the main article within 1$\%$, all reported calculations are obtained numerically by introducing a cutoff $-30\le n\le 30$ in the infinite sum Eq.~(\ref{eq:eigen}).

\subsection{Band structures in corrugated graphene systems}

To study the electronic structures in our periodic corrugated graphene system, we firstly model a single corrugation locating at $-{\rm w} \le x < {\rm w}$ using the hyperbolic-tangent function
\begin{align}
h(x) = \frac{D}{2} ({\rm ArcTan}[\frac{1}{b}(x+\frac{{\rm w}}{2})]- {\rm ArcTan}[\frac{1}{b}(x-\frac{\rm w}{2})]),
\end{align}
where $D = 8$~nm, w$= 100$~nm, and $b=11.5$ nm are the height, the width, and the slope of a corrugation, respectively. We then repeat the $h(x)$ with a period 2w and insert it into Eqs. (\ref{eq_a0}), (\ref{eq_a3}) to solve both the pseudo-gauge fields. The intralayer pseudo-magnetic fields in different valleys are also obtained as the curl of $\mathbf{A}_0$ (see the bottom panel in Fig.~1b in the main article).
 
With the knowledge of both pseudo-gauge fields, the Hamiltonian in Eq.~(\ref{h_2band}) can be diagonalized with the basis in Eq.~(\ref{eq:basis}). The calculated band structures of our corrugated BLG for $\phi=7.7^{\circ}$ without and with the skew interlayer coupling are shown in Figs.~4a \& b and Figs.~4d \& e in the main article, respectively. For MLG with the same periodic corrugation, the effective Hamiltonian is different\cite{car2016}. Following the same steps from Eq.~(\ref{h_2band}) to Eq.~(\ref{eq:basis}) but with the MLG Hamiltonian and the eigenbasis, the band structure of the MLG corrugation can also be solved. The calculated band structure of MLG corrugation is shown in Extended Data Fig.~\ref{Extended5}, in which the pseudo-magnetic states as well exhibit a Rashba-like valley-orbit splitting along the $k_y$ axis. The slope of bands near zero energy is linear due to the massless nature of the low-energy electrons in MLG. In contrast, the corrugated BLG has flat bands near zero energy (Figs.~4a, b in the main article) because of the onsite interlayer coupling\cite{mcc2013} and is similar to those observed in magic-angle graphene superlattices\cite{cao_nature18a,cao_nature18b}.

\subsection{Calculations of transport properties and angle of resistivity anisotropy}
With the calculated band structure of the corrugated BLG, the conductivity tensor can be computed by using the linear-response function ($\hbar = 1$)
\begin{align}
\sigma_{ab}=\frac{-e^2 \tau}{1+i \omega\tau}\int \frac{d^2 \mathbf{k}}{(2\pi)^2} \delta(\epsilon(\mathbf{k})-\epsilon_{\text{F}}) \partial_{a} \epsilon(\mathbf{k}) \partial_{b} \epsilon(\mathbf{k}),
\label{eq:conduct}
\end{align} 
with $\partial_a = \partial/\partial_{k_a}$ and $\tau$ is the relaxation time. There are two mechanisms to smooth the calculated conductivity for the discrete delta function $\delta(\epsilon(\mathbf{k})-\epsilon_{\text{F}})$. The first is the finite temperature smearing that can be taken into account simply by replacing the delta function with the Fermi-Dirac distribution, while the second is due to the scattering centers naturally existing in materials\cite{shi2002}. Here we include the second mechanism by effectively increasing the temperature as $T+T_{\rm eff}$ with $T_{\rm eff} \approx 20$~K in the Fermi-Dirac distribution. We have numerically verified that the values of conductivity modulated by $T_{\rm eff}$ are approximately the same as those considering a Lorentzian broadening\cite{shi2002} with a width $k_B T_{\rm eff}$.

The resistivity tensor can be obtained by inverting the conductivity tensor calculated using Eq.~(\ref{eq:conduct}). The relaxation time of $\tau=1.5\times10^{-13}$~s is used to give rise to consistent results with the experiments. The off-diagonal resistivity $\rho_{yx}$ for our corrugated BLG is plotted in Figs.~3b,d in the main article. In addition, we also estimate the $\rho_{yx}$ for the same system with different crystal orientation. When the angle $\phi$ between the graphene zigzag axis and the corrugation direction increases, the pseudo-magnetic field increases and consequently the pseudo-PHE arises. The reason that the pseudo-PHE behave differently near the NP is that the effects of skew interlayer coupling are generally very different for electron transport near and away from the NP\cite{cse2007}. For corrugated MLG with the band structure calculated using the same technique, however, we find that $\rho_{xy} = \rho_{yx} = 0$ since the band dispersions along the $k_x$ axis are symmetric. The pseudo-magentic-field induced PHEs are unique phenomena only in corrugated BLG system.

To further understand the nature of this pseudo PHE, we calculate the angular variation of resistance anisotropy as a function of the Fermi energy from both experimental observations and theoretical results. The obtained resistivity matrix can be rotated back to be a diagonal one by applying the rotation matrix $r(\theta)=\{\{\cos\theta,-\sin\theta\},\{\sin\theta, \cos\theta\}\}$
\begin{align}
r^+(-\theta)\left (
  \begin{array}{cc}
    \rho_{xx} &\rho_{xy}\\
 \rho_{yx} & \rho_{yy}  \\
  \end{array}
\right )r(-\theta) 
= \left(\begin{array}{cc}
    \rho_{x' }&0\\
 0 & \rho_{y'}  \\
  \end{array}
\right ),
\end{align}
where $\rho_{xy}=\rho_{yx}$ in our BLG system with time-reversal symmetry.
The resistivity $\rho_{x'}$ and $\rho_{y'}$ are the resistivities along the principal x'- and y'-axes, respectively, and so that the above matrix has no transverse (Hall-like) response.  The resistivity-anisotropy angle $\theta$ can be estimated as $\theta=0.5\arctan\frac{2\rho_{xy}}{\rho_{yy}-\rho_{xx}}$. 
For a conventional crystalline system injected by a misaligned current, the angle $\theta$ is a constant angle between the current direction and crystal axes, 
and thus is independent to the Fermi energy. In our corrugated BLG, the $\theta$ obtained from both experimental and theoretical results are shown in Fig.~3e. 
This angle, however, changes dramatically with the Fermi energy variation and even switches its sign. The observed resistivity anisotropy is unconventional and thus indicates a new type of Hall-like response in corrugated BLG system. 

\subsection{Calculations of Berry curvature and Berry curvature dipole}
With the solved eigenstates of the corrugated BLG, the Berry curvature ${\bf \Omega}^n$ of a $n-$th band at momentum ${\bf k}$ can be estimated. 
In two-dimensional materials, only the out-of-plane component of the Berry curvature is nonzero, and it is a summation of contributions over the eigenstates\cite{xiao_rmp10}: 
\begin{align}
\Omega^n_{z}({\bf k})=i\sum_{n'\neq n}\frac{\langle n|\partial_x \hat{H}|n'\rangle\langle n'|\partial_y \hat{H}|n\rangle -(x\leftrightarrow y)}{(\epsilon_n-\epsilon_{n'})^2},
\label{eq:berrycurvatue}
\end{align}
with $\epsilon_n$ being the eigenstate energy. 
To open the band crossings to obtain the Berry curvature\cite{sodemann_prl15}, 
we add a small asymmetry profile $h_a=-0.2D \cos\big[\pi(x-{\rm w}/2)/{\rm w}\big]-0.1D \cos\big[2\pi(x-{\rm w}/2)/{\rm w}\big]$ that is an order of magnitude smaller than the Eq.(7).
For the tilted mini-Dirac cone shown in Fig.~1c in the main article, the calculated Berry curvature is provided in Extended Data Fig.~2.

The Berry curvature dipole can be further estimated from the solved Berry curvature. For the AC field is applied perpendicular to the corrugations (x direction), the nonlinear Hall response appearing along the corrugation (y direction) is caused by the Berry curvature dipole  
\begin{align}
D_x(E_F,T) = \sum_n\int_k f_0(\epsilon_n-E_F,T) (\partial_x\Omega^n_z),
\end{align}
where $f_0$ indicates the Fermi-Dirac distribution, and $E_F$ is the Fermi energy. The calculated Berry curvature dipole is provided in the main article.

\subsection{Obtaining Berry curvature dipole from measured data}
In this section, we will show how to obtain the experimental data of Berry curvature dipole by using informations of the non-linear Hall voltage, the linear-response resistivities, and the amplitude of applied AC current. When the AC current is injected along the x direction (perpendicular to the corrugations), the induced nonlinear Hall current along the y direction is\cite{sodemann_prl15}
\begin{align}
j^{2\omega}_y=\frac{1}{2\hbar^2}\frac{e^3\tau}{1+i\omega\tau}D_x (\varepsilon^\omega_x)^2,
\end{align}
with $\varepsilon^\omega_x$ is the amplitude of external AC electric field driven by the injected AC current.
By considering the low AC frequency $\omega\tau \ll 1$, the Berry curvature dipole $D_x$ can be approximated as
\begin{align}
D_x\approx \frac{2\hbar^2}{e^3\tau}\frac{j^{2\omega}_y}{(\varepsilon^\omega_x)^{2}}= \frac{2\hbar^2}{e^3\tau}L(\rho^{\omega}_y)^{-1}\frac{V^{2\omega}_y}{(I_x^\omega\rho_x^\omega)^2},
\end{align}  
with $L$ is the width of the Hall bar, $I^{\omega}_x$ is the amplitude of applied AC current, $V^{2\omega}_y$ is the nonlinear Hall voltage. We note that it is necessary to take into account the different resistivities in x and y directions in the above equation, since the transport in our corrugated BLG system is very anisotropic. 

\subsection{Calculations of transverse conductivity in an effective model}

Two key ingredients resulting in our pseudo-PHE are skew interlayer coupling and pseudo-magnetic field. When the graphene is curved on a corrugated substrate, the local lattice extension drives the vector potential for electrons and results in the pseudo-magnetic field. We then estimate the linear Hall response observed in corrugated BLG by an effective model by using Einstein relation and the classical linear response theory \cite{mec2005, cha2017}
\begin{align}
\sigma_{xy}=e^2 N\int_0^\infty \langle v_x(t)v_y(0)\rangle e^{-t/\tau}dt
\end{align}
where x and y is the direction across and along the corrugations, respectively, N is the density of states, and  $\tau$ is the relaxation time. The brackets denote an average over the available phase space. We consider a flank of a single corrugation that provides ${\bf B}_{\rm eff} (x)$ shown in Fig.~1b.  The velocity correlation function is obtained by solving the Newton equation of motion $m^\star \dot{{\bf v}}=-e {\bf v}\times {\bf B}_{\rm eff}(x)$ with $m^\star$ is the carrier effective mass. The theoretical approaches in above two equations have been widely used to simulate the magnetoresistance in two dimensional electron gases (2DEGs) subject to inhomogeneous magnetic field \cite{sch2015}. The key difference between 2DEGs and bilayer graphene is the carrier velocity distribution. Carriers in 2DEGs are simulated to initially move in random directions with a fixed Fermi velocity. In bilayer graphene system, however, we simulate that carrier initially move randomly with the direction-dependent velocity $v_i = \partial \epsilon(p)/\partial p_i$ with $i= \{x,y\}$ and $\epsilon(p)$ is the warped energy dispersion of K valley in BLG (Equation (41) in Ref. \cite{mcc2013}).

To obtain the conductivity tensor, we have averaged the velocity correlation function with respect to 2000000 sampling points in the phase space, thereby reaching an accuracy within 0.001$\%$. The experimental parameters are adopted in the calculations, where ${\bf B}_{\rm eff} (x)$ having a periodic wavelength 100 nm and  estimated by multiplying the mean free path equal to 150 nm and the approximated electron velocity in graphene $v_F = 10^6$m/s. The obtained Hall conductivity is nonzero, much smaller than the longitudinal conductivity $\sigma_{xx}$, and increases almost linearly with the Fermi energy. All of these properties consistent with our previous theoretical results based on solving the band structure. Since the Hall response is driven by the interplay between  ${\bf E}_{\rm eff} (x)$  and  ${\bf B}_{\rm eff}$, this effect is the same in different valley due to the time-reversal symmetry reversing signs of both effective fields.

We have verified that the pseudo PHE exists in the effective model. We further checked that the role of the periodicity by testing two systems. We firstly reduce the mean free path from 150 nm to 50 nm. In such a short mean free path, the electron can only propagate around 50 nm that is much shorter than a period of  ${\bf B}_{\rm eff} (x)$. In addition, we also estimate the conductivity in a system with only one single flank of the corrugation. The remarkable Hall conductivity exits in both tested systems.

\subsection{Data availability.} The data that support the plots within this paper and other findings of this study are available from the corresponding author on reasonable request.

\subsection{Code availability.} The code that supports the theoretical plots within this paper is available from the corresponding author upon reasonable request.

%\bibliography{scibib}

%insert Fig.1 %
\begin{figure} [b!]
  \centering 
  \includegraphics[width=0.95\columnwidth]{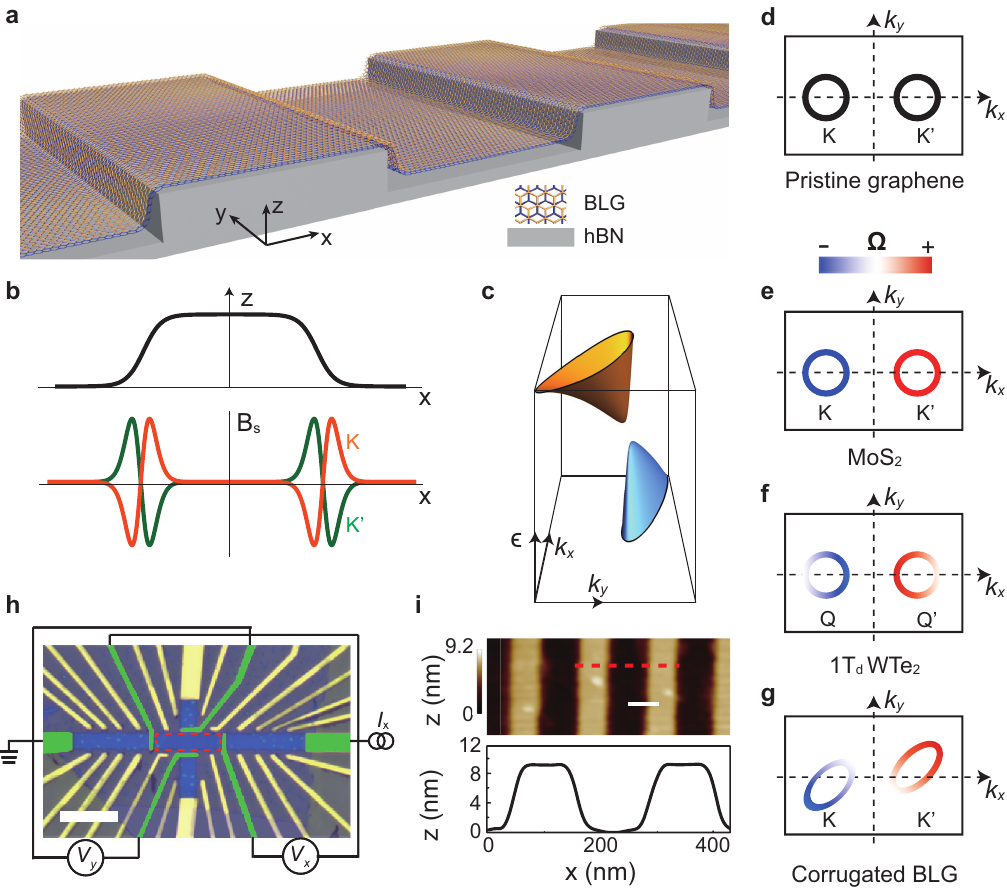}
  \caption{\textbf{Artificially corrugated bilayer graphene (BLG) and its electronic properties.} \textbf{a}, Schematic of our artificially corrugated BLG device. BLG is mechanically deformed into a corrugated configuration defined by the hBN underneath. \textbf{b}, Illustration of the BLG corrugation profile and its corresponding strain-induced pseudo-magnetic field $B_{\text{S}}$ distribution along the $x$-direction. $B_{\text{S}}$ only occurs around the pattern steps that are far enough apart to isolate each of the antiferromagnetic pair of $B_{\text{S}}$, which on the other hand have opposite sign for the K and K$'$ valley. \textbf{c}, Calculated band structure of corrugated BLG, showing one of the titled mini-Dirac cones with non-trivial energy dispersion. This is a zoom-in plot of Fig.~4f. }
  \label{figure1}
\end{figure}
\addtocounter{figure}{-1}
\begin{figure} [t!]
  \caption{(Continued) \textbf{d-g}, Classification of 2D materials via Fermi circle and Berry curvature $\mathbf{\Omega}$. For pristine graphene (\textbf{d}), the Fermi circle (Dirac cone) is of perfect circular symmetry and $\mathbf{\Omega} = 0$ over the whole Brillouin zone as a consequence of the time reversal and inversion symmetry. The gapped graphene and most of the TMDCs, e.g. MoS$_2$ (\textbf{e}) also possess a trivial Fermi circle but have nonzero Berry curvature as a result of broken inversion symmetry. For materials with very low crystalline symmetry, e.g. 1T$_d$~WTe$_2$ (\textbf{f}), a titled Dirac cone with Berry curvature dipole (which can be viewed as non-uniform Berry curvature distribution on the Fermi circle) will form\cite{xu_np18}. In contrast, our corrugated BLG (\textbf{g}) possesses a titled Dirac cone with nontrivial valley-orbit coupled band dispersion, as well as the Berry curvature dipole. \textbf{h}, Optical image of the corrugated BLG device with measurement circuit. Metallic electrodes highlighted in green represent the contacts used for Hall resistivity $\rho_{yx}$ measurements. The red dashed rectangle highlights the area of BLG corrugation. Scale bar (white): 10~$\mu$m. \textbf{i}, AFM topography image showing the BLG corrugation after the BLG is transferred onto the hBN. Transport measurements are performed in different devices with the same geometry design, and all of the measured devices show consistent results. The red dashed line on the top panel corresponds to the position of the height profile shown in the bottom panel. Scale bar (top panel, white): $100$~nm.}%missing
\end{figure}

%insert Fig.2 %
\begin{figure}
  \centering 
\includegraphics[width=1\columnwidth]{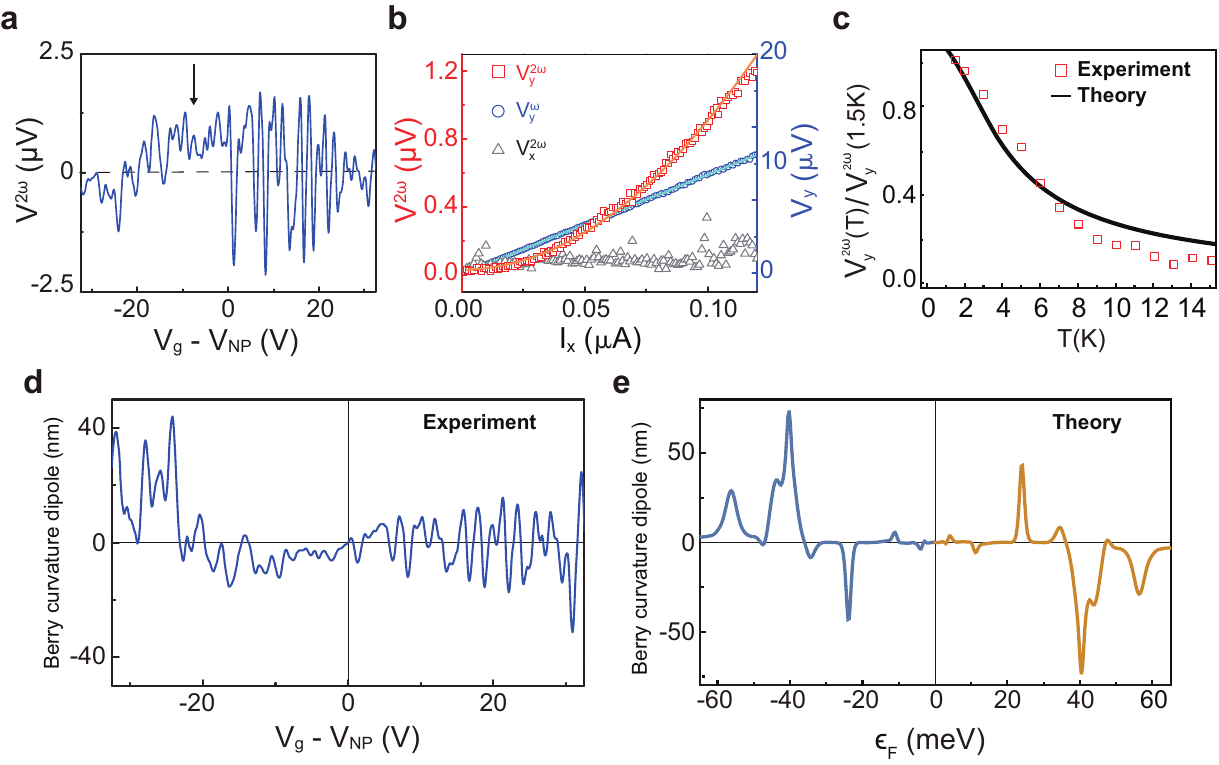}
\caption{\small \textbf{Nonlinear anomalous Hall effect (AHE) and Berry curvature dipole.} \textbf{a}, Nonlinear AHE --  the second-harmonic Hall voltage response $V_y^{2\omega}$ -- measured as a function of $V_g$. \textbf{b}, The first-harmonic Hall voltage ($V_y^{\omega}$, blue circle), second-harmonic Hall voltage ($V_y^{2\omega}$, red square), and second-harmonic longitudinal voltage ($V_x^{2\omega}$, grey triangle) measured as a function of the driving current magnitude $I_x$, for $V_g-V_{\text{NP}} =-8$~V (black arrow in \textbf{a}) and at $T = 5$~K. The first- and second-harmonic Hall voltage are well fitted by a quadratic and linear function shown in light blue and orange curves. The linear and quadratic I-V characteristics for the pseudo-PHE and nonlinear AHE are consistently observed at various $V_g$. }
  \label{figure2}
\end{figure}
\addtocounter{figure}{-1}
\begin{figure} [t!]
  \caption{(Continued) \textbf{c}, Temperature dependence of the experimentally measured (red square) and theoretically calculated nonlinear AHE. In order to have a direct comparison between experiment and theory, the nonlinear Hall voltage is normalized by the signal at the lowest temperature of $1.5$~K. Our theoretical calculations consider only the temperature effect on the Berry curvature dipole and assume the transport relaxation time to be constant, which may be responsible for the deviations between the measured and calculated values at high temperature. Consistent results have been observed at various $V_g$. \textbf{d}, Berry curvature dipole as a function of back gate voltage, derived from the experimentally measured nonlinear AHE voltage (\textbf{a}) and other transport parameters (see Methods for detailed derivation). \textbf{e}, Same as \textbf{d} but obtained from the theoretical calculation based on the band structure predicted for our corrugated BLG system.}
\end{figure}

\makeatletter
\@fpsep\textheight
\makeatother

%insert Fig.3 %
\begin{figure} 
  \centering 
  \includegraphics[width=0.78\columnwidth]{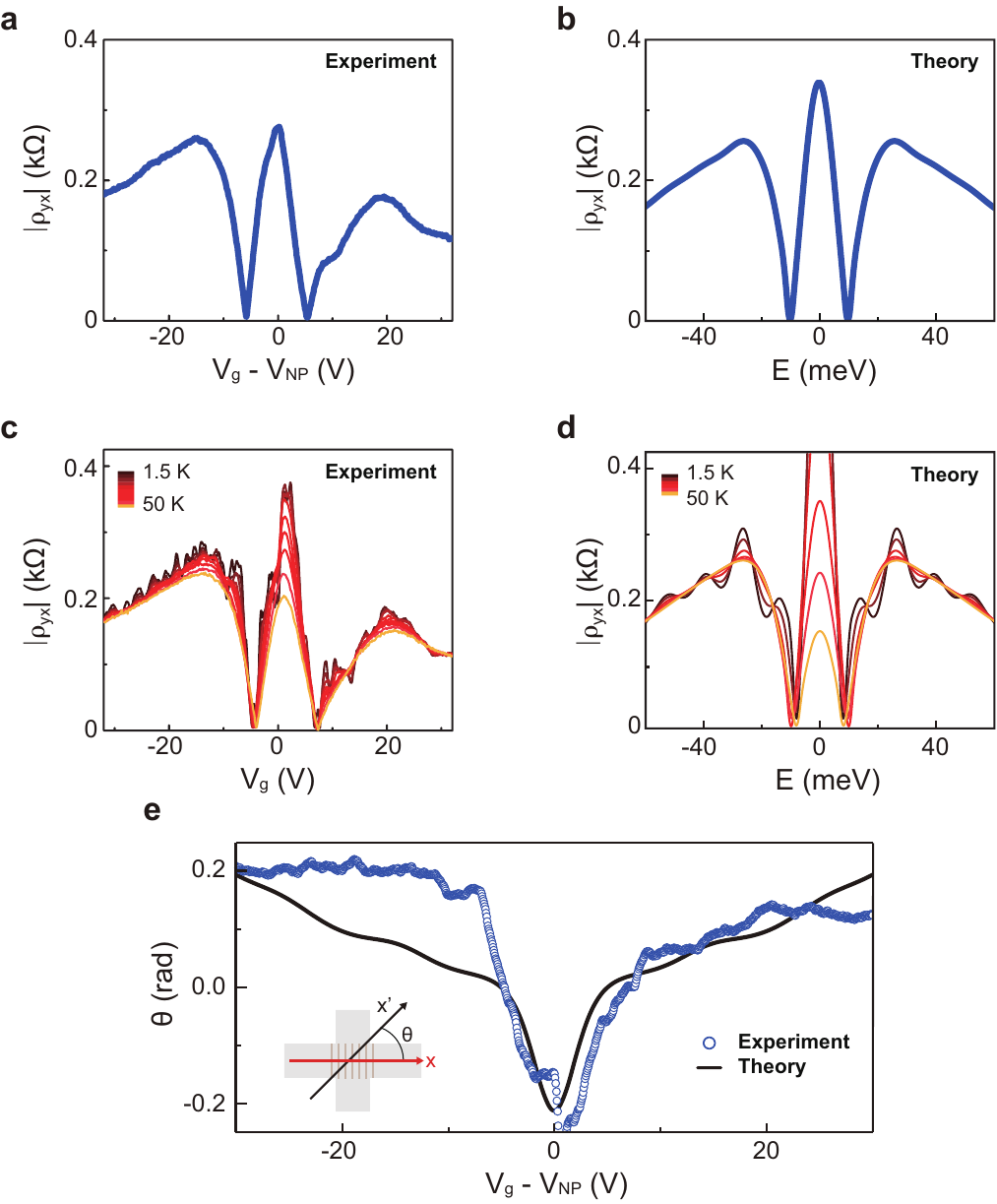}
  \caption{\small \textbf{Pseudo-magnetic-field induced planar Hall effect (pseudo-PHE) and pseudo-magnetoresistance anisotropy.} \textbf{a,b}, Magnitude of the experimentally measured (\textbf{a}) and theoretically calculated (\textbf{b}) planar Hall resistivity $|\rho_{yx}|$ as a function of the back gate voltage $V_g$ and Fermi level position $\epsilon_\text{F}$ in the corrugated BLG at $T=30$~K, respectively. Experimental results of the pseudo-PHE resistivity $\rho_{yx} = V_y/I_x$ are obtained by measuring the linear-response Hall voltage $V_y$ transverse to the applied current $I_x$ as illustrated in Fig.~1h. For BLG the lever arm\cite{lee_science14} for converting the gate voltage to energy ($\alpha=eV_g/\epsilon_{\text{F}}$) is about $0.5 \times 10^3$. } 
  %\cite{lee_science14}
  \label{figure3}
\end{figure}
\addtocounter{figure}{-1}
\begin{figure} [t!]
  \caption{(Continued) \textbf{c,d}, Same as \textbf{a,b} but for various temperatures ranging from $1.5$ to $50$~K. \textbf{e}, The angular variation of the anisotropy axis of pseudo-magnetoresistance as a function of the back gate voltage. The angle $\theta$ is defined between the principle axes of the anisotropic resistance $x'$ and the direction perpendicular to the corrugation ($x$ direction), as shown in the inset.}%missing
\end{figure}

%insert Fig.4 %
\begin{figure}
  \centering 
\includegraphics[width=1\columnwidth]{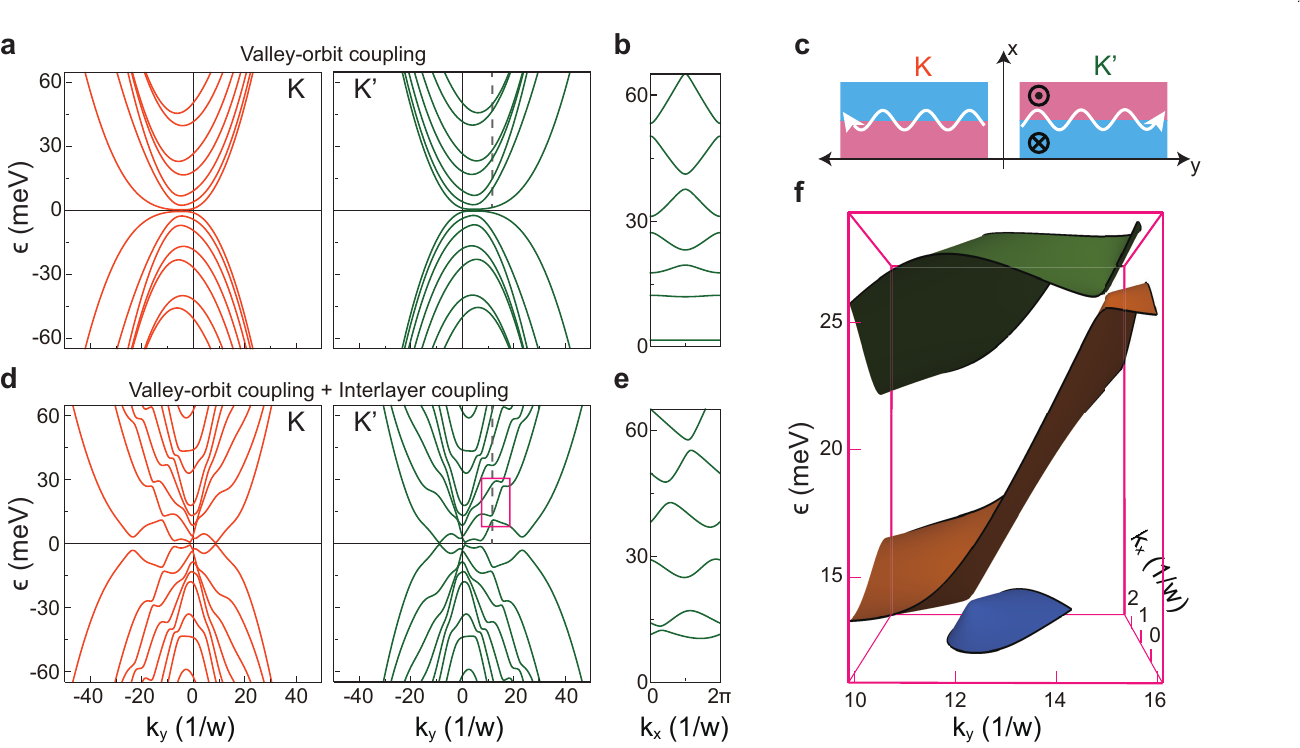}
\caption{\small \textbf{Electronic band structure of corrugated BLG.} \textbf{a}, Calculated band dispersion of the corrugated BLG without considering the skew interlayer hopping, for the K (orange) and K$'$ (green) valleys along the $k_y$ direction. The pseudo-magnetic states exhibit a Rashba-like valley-orbit splitting. \textbf{b}, The band dispersion along the $k_x$ direction at $k_y = 12$ (corresponding to the dashed line in \textbf{a}) shows a conventional symmetric Brillouin zone. \textbf{c}, Illustration of the low-energy snake-like states that occur around the corrugation steps. Electrons in the K and K$'$ valley experience the effective Lorentz force with opposite directions as $B_\text{S}$ is reversed, and consequently move towards the opposite directions. \textbf{d,e}, Same as \textbf{a,b} but with the skew interlayer hopping taken into account. Such a interlayer hybridization does not merely mix but more importantly warp these pseudo-LL magnetic states, resulting in tilted mini-Dirac cones in which nonzero Berry curvature dipoles are expected. \textbf{f}, The calculated band dispersion along both $k_x$ and $k_y$ (zoom-in on the region highlighted by the rectangle in \textbf{d}) shows the titled mini-Dirac cones with highly distorted energy dispersion.}
\end{figure}

\renewcommand{\figurename}{Extended Data Figure}
\setcounter{figure}{0}

%insert Fig. S1 %
\begin{figure}[h]
  \centering 
\includegraphics[width=0.6\columnwidth]{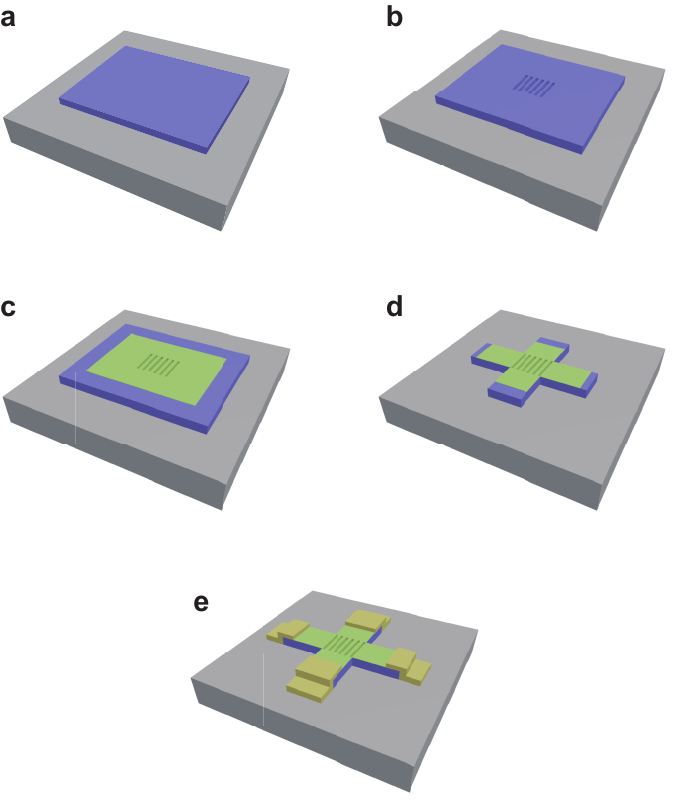}
\caption{\textbf{Fabrication processes of our corrugated BLG.} \textbf{a-e}, Schematic illustrations showing a series of our fabrication process steps. The colour representation: the SiO$_{2}$ substrate is highlighted in grey, hBN in purple, BLG in green, and the Cr/Au contacts in yellow. Fabrication details for each step are described in Materials and Methods.}
\label{Extended2}
\end{figure}

\vspace*{1cm}
%insert Fig. S2 %
\begin{figure}[h]
  \centering 
\includegraphics[width=0.7\columnwidth]{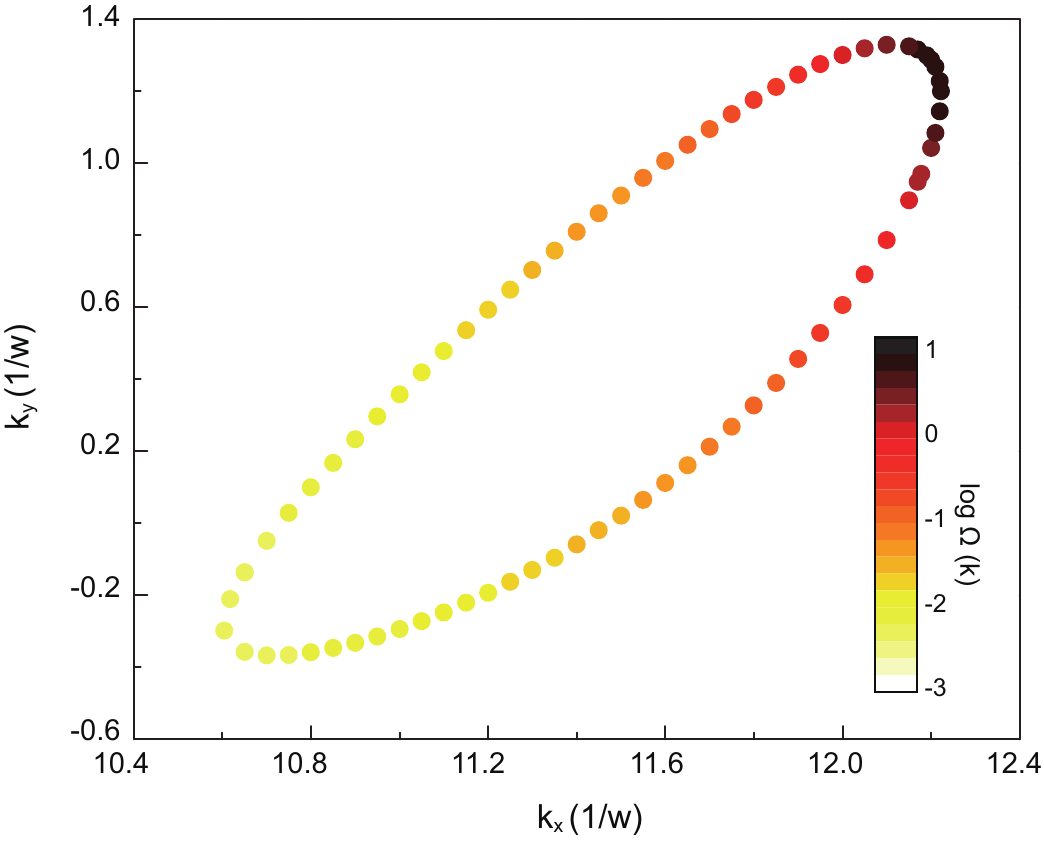}
\caption{\textbf{Berry curvature dipole of corrugated BLG.} Calculated Berry curvature of the tilted mini-Dirac cone shown in Fig.~1c in the main article. Large scale inhomogeneity (for more than three orders of magnitude) of the Berry curvature distribution around the warped Fermi circle is clearly observed in such a BLG corrugation. The Fermi energy is set at $13.25$~meV.}
\end{figure}

\vspace*{1cm}
%insert Fig. S3 %
\begin{figure}[h]
  \centering 
\includegraphics[width=0.9\columnwidth]{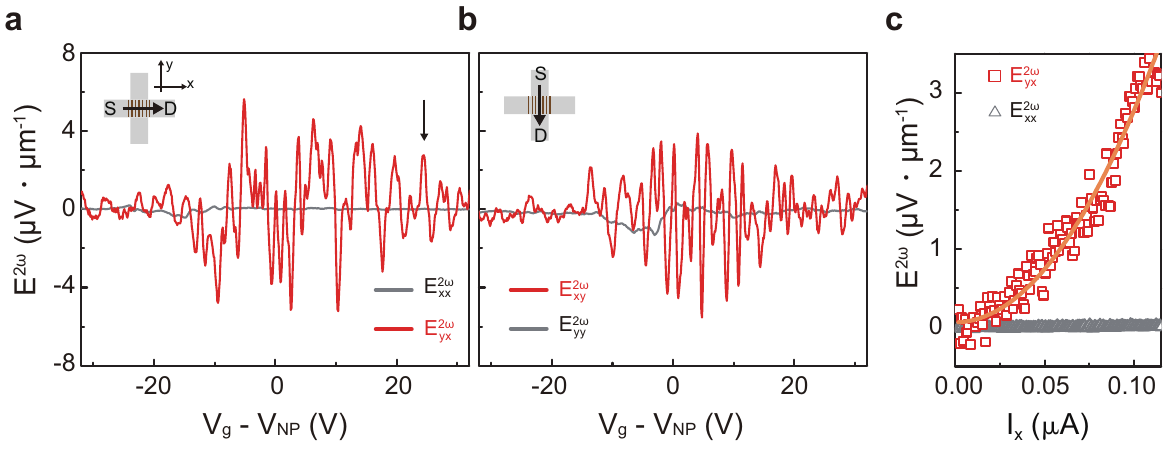}
\caption{\small \textbf{90-degree Hall angle and reproducibility of nonlinear AHE} The second-harmonic transverse ($E_y^{2\omega}$, grey) and longitudinal ($E_x^{2\omega}$, red) electric field measured as a function of gate voltage $V_g$ when the current is applied (\textbf{a}) perpendicular to and (\textbf{b}) parallel to the corrugation direction at $T= 5$~K. (\textbf{c}) The current-voltage characteristic of $E_x^{2\omega}$ and $E_y^{2\omega}$ for $V_g - V_{NP} = 20.5$~V (black arrow in (a)). The second-harmonic Hall voltage are well fitted by a quadratic function shown in orange curve. The quadratic I-V characteristics for the nonlinear AHE are consistently observed at various $V_g$.}
\end{figure}

\vspace*{1cm}
%insert Fig. S4 %
\begin{figure}[h]
  \centering 
\includegraphics[width=0.7\columnwidth]{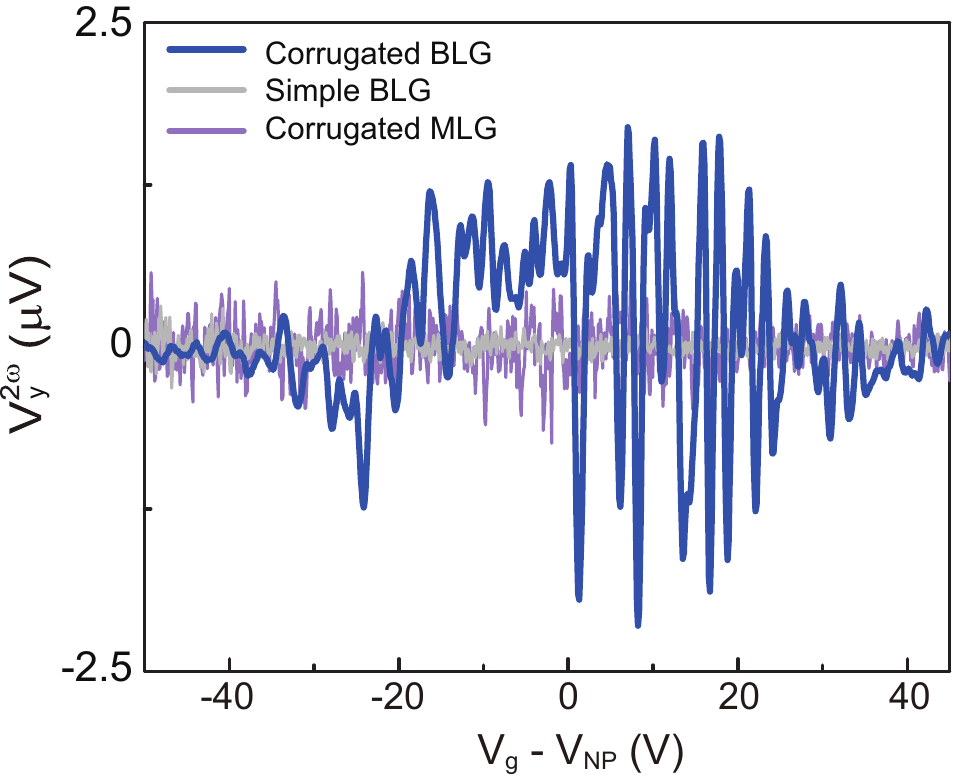}
\caption{\small \textbf{Comparison of nonlinear AHE responses in corrugated BLG, simple BLG, and corrugated MLG.} The second harmonic Hall voltage response $V_y^{2\omega}$ as a function of the back gate voltage is measured using standard lock-in technique in the corrugated BLG, simple BLG and corrugated MLG for comparison. The signal in the corrugated BLG is well reproducible with the phases of the second-harmonic responses being locked at $\pm90^{\circ}$ with respect to the driving reference, which is in agreement with the expectations for the nonlinear measurements. In contrast, the small, fluctuant $V_y^{2\omega}$ in simple BLG and corrugated MLG are noise signal because they are irreproducible and the phase cannot be locked, showing the absence of nonlinear AHE in these two systems.}
\end{figure}

\clearpage

%inset Fig. S5 %
\begin{figure}[b!]
  \centering 
\includegraphics[width=0.72\columnwidth]{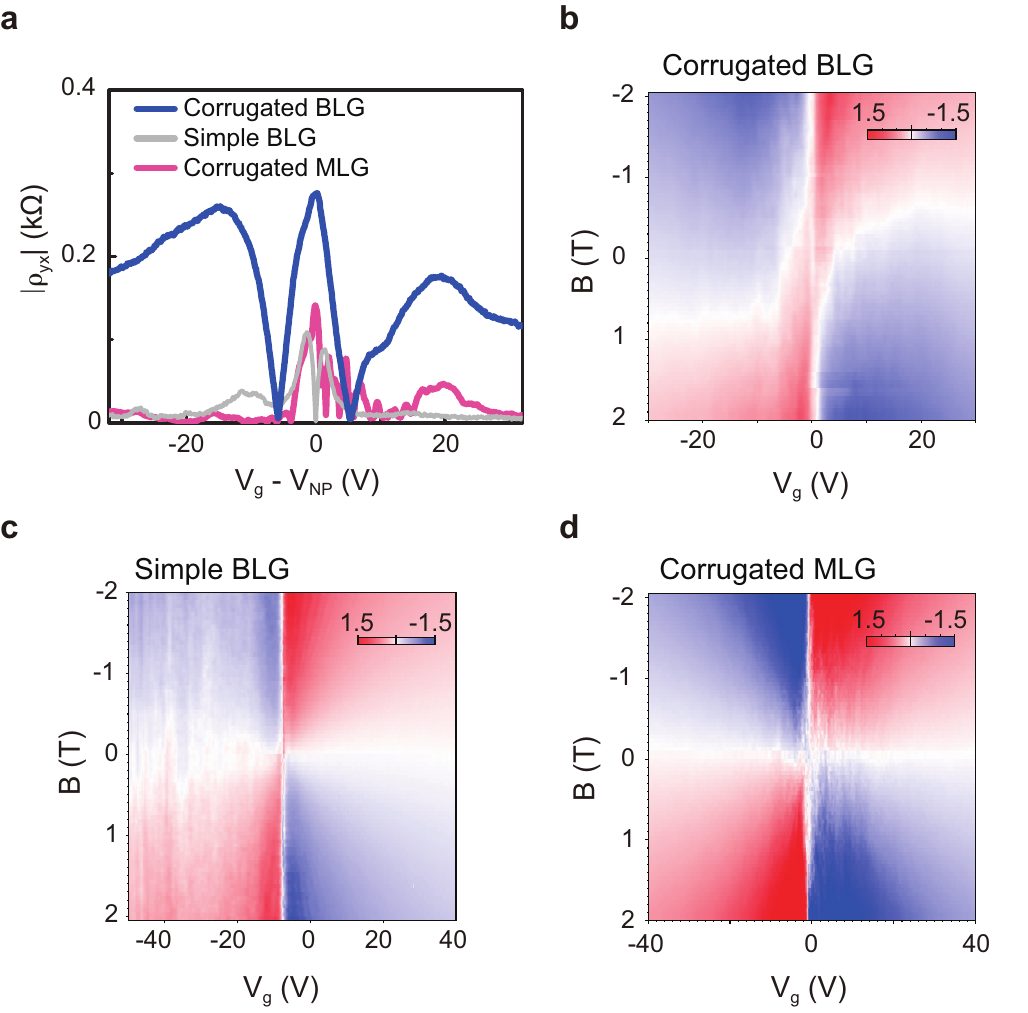}
\caption{\small \textbf{Comparison of linear transverse responses in corrugated BLG, simple BLG, and corrugated MLG.} \textbf{a}, The linear transverse resistivity $|\rho_{yx}|$ as a function of the back gate voltage is measured in the corrugated BLG, simple BLG and corrugated MLG. In comparison with the $|\rho_{yx}|$ in corrugated BLG, the signal is negligibly small for simple BLG and corrugated MLG except near the charge neutrality point. The transverse response near the NP may be due to the charge inhomogeneity because of the electron-hole puddles or the imperfection of the Hall bar that were commonly observed in graphene near the neutrality point\cite{amado_njp10,caridad_ncomm16}, and is not the major focus of this work. It is the transverse response that can be seen over a very wide range of Fermi energy and well explained by the model of pseudo-PHE that interests and lies at the heart of this work.}
%  \label{extended figure3}
\end{figure}
\addtocounter{figure}{-1}
\begin{figure} [t!]
 \caption{(Continued) \textbf{b-d}, Colourmaps of the transverse resistivity $\rho_{yx}$ as a function of $V_g$ and $B$ in the corrugated BLG (\textbf{b}), simple BLG (\textbf{c}) and corrugated MLG (\textbf{d}). In corrugated BLG, interplay between the zero-magnetic-field pseudo-PHE and the classical Hall effect can be clearly seen, exhibiting an anti-crossing structure in the colourmap. In contrast, for the simple BLG and corrugated MLG, a standard colour plot of the Hall effect in graphene with four quadrants crossing at zero magnetic field and the neutrality point is shown. 
}
\end{figure}

\clearpage

\vspace*{1cm}
%insert Fig. S6 %
\begin{figure}[h]
\begin{center}
\includegraphics[width=0.9\columnwidth]{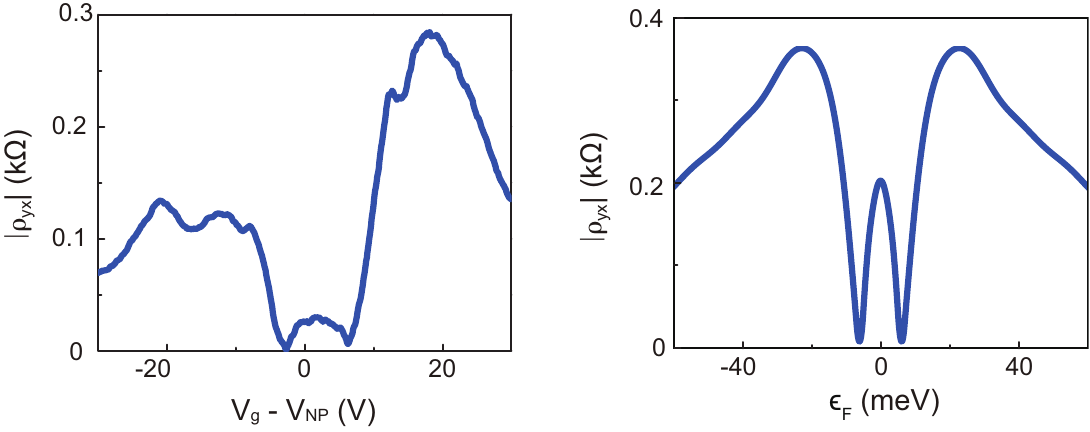}
\end{center}
\caption{\small \textbf{Reproducibility of pseudo-PHE} \textbf{a}, The linear Hall resistivity $|\rho_{yx}|$ measured as a function of gate voltage $V_g$ in another corrugated BLG device. \textbf{b}, The linear Hall resistivity that is theoretically calculated as a function of Fermi level position $\epsilon_\text{F}$ in the corrugated BLG with $\phi = 10^{\circ}$, where $\phi$ is the angle between the graphene zigzag axis and the corrugation direction.}
\end{figure}

\vspace*{1cm}
%insert Fig. S7 %
\begin{figure}[h]
\begin{center}
\includegraphics[width=1\columnwidth]{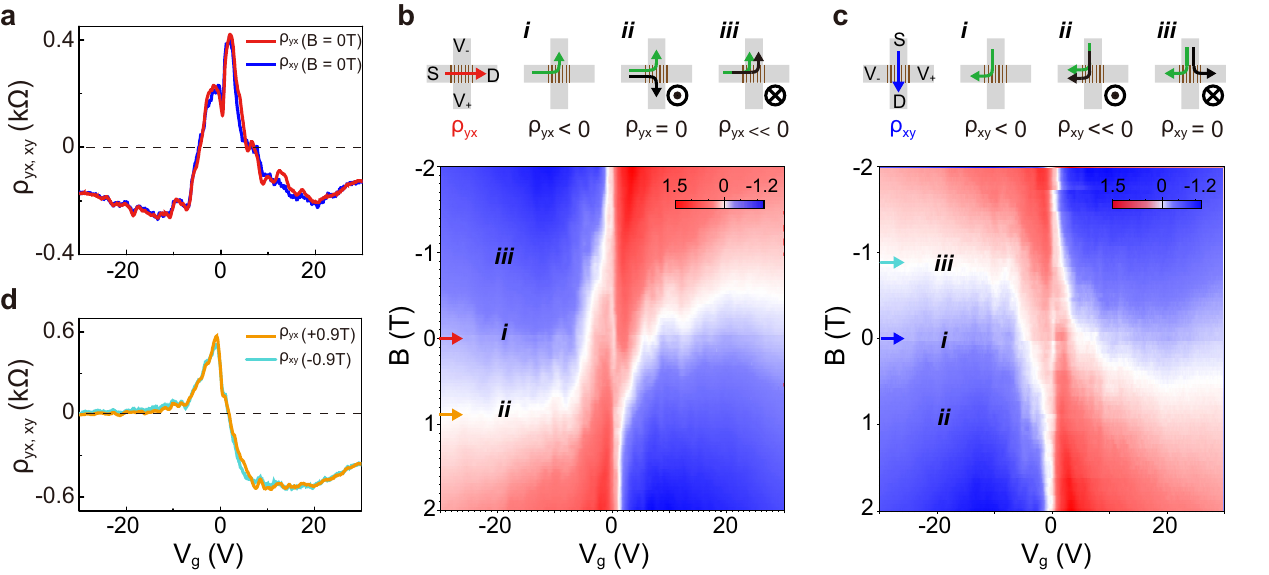}
\end{center}
\caption{\small \textbf{Time reversal invariant pseudo-PHE.} \textbf{a}, The planar Hall resistivities $\rho_{yx}$ (see top insets in \textbf{b} for measurement setup) and $\rho_{xy}$ (see \textbf{c}, top insets) measured as a function of $V_g$ at zero external magnetic field $B=0$. Although the current directions and measurement setups for $\rho_{yx}$ and $\rho_{xy}$ are very different with respect to the corrugation and device geometry, the fact the $\rho_{xy} = \rho_{yx}$ indicates that the resistivity is a symmetric tensor and the observed pseudo-PHE is indeed time reversal symmetric. \textbf{b, c}, Colour rendition of $\rho_{yx}(B,V_g)$ (\textbf{b}) and $\rho_{xy}(B,V_g)$ (\textbf{c}). The top insets show representative current trajectories for each region labelled in the colour plot. Interplay between the pseudo-PHE and the HE results in an asymmetric overall Hall resistivity, e.g., the Hall resistivity $\rho_{yx} (B=+0.9~\text{T})$ (region labelled \textbf{ii} in \textbf{b}) is completely different to $\rho_{yx} (B=-0.9~\text{T})$ (region labelled \textbf{iii} in \textbf{b}). However, $\rho_{yx} (B=+0.9~\text{T})$ (region labelled \textbf{ii} in \textbf{b}) is nearly the same as $\rho_{yx} (B=+0.9~\text{T})$ (region labelled \textbf{iii} in \textbf{c}). \textbf{d}, $\rho_{yx} (B=+0.9~\text{T})$ and $\rho_{yx} (B=+0.9~\text{T})$ are explicitly plotted as a function of $V_g$ to demonstrate the equivalence. }
\label{Fig3s}
\end{figure}

\clearpage

\vspace*{1cm}
%insert Fig. S8 %
\begin{figure}[h]
  \centering 
\includegraphics[width=0.9\columnwidth]{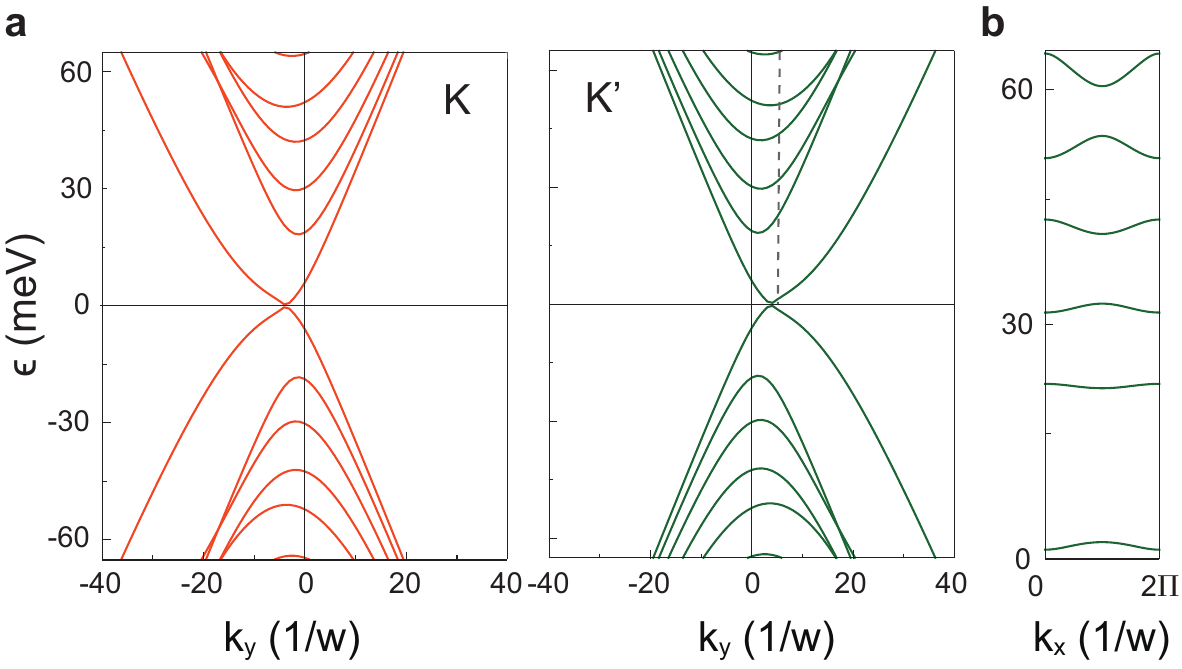}
\caption{\small \textbf{Electronic band structure of corrugated MLG.} \textbf{a}, Calculated band dispersion of the corrugated monolayer graphene for K (orange) and K$'$ (green) valley along $k_y$ direction at $k_x = 0$. \textbf{b}, The band dispersion along $k_x$ direction at $k_y = 5$ (corresponding to dashed line in \textbf{a}) shows a conventional symmetric Brillouin zone.}
\label{Extended5}
\end{figure}

\end{document}